\documentclass[prb,twocolumn,showpacs,preprintnumbers,amsmath,amssymb]{revtex4}
\usepackage{bm}
\usepackage{amsmath}
\usepackage{amsthm}
\usepackage{amssymb}
\usepackage{graphicx}

\newcommand{\eref}[1]{(\ref{#1})}

\begin{document}

\title{Nonlinear Dynamics of Ultra-Cold Gas:
Collapse of Bose Gas With Attractive Interaction}
\author{V.V.  Flambaum}
\affiliation{Institute for Nuclear Physics, 630090 Novosibirsk, Russia. Current address:School of Physics, University of New South Wales, Sydney 2052, Australia }
\author{E. Kuznetsov}
\affiliation{Institute of Automation and Electrometry, 630090 Novosibirsk, Russia}
\date{\today}

\begin{abstract}
Solutions for  the Nonlinear Schr\"{o}dinger equation for collapsing Bose gas with attraction, $ H = \int \left[\frac{\hbar^2}{2m}|\nabla \psi|^2+\frac{G}{2}|\psi|^4+\frac{\beta}{3}|\psi|^6\right]dr.$
This is a copy of the paper published in 1992 in Proceedings of NATO Advanced Research workshop on Singularities in Fluids, Plasmas and Optics (Heraklion, Greece) edited by R.E. Caflisch and G.C. Papanicolaou  (Kluwer Academic). 
\end{abstract}


\maketitle  

\section{Introduction}
In recent years it has become possible to cool and to trap neutral atoms using the resonant laser radiation (see, for example, \cite{Monroe90} and references therein). The main idea of cooling and trapping is connected with the dependence of radiation pressure taking into account the resonant absorption and photon radiation by atoms (see for example, \cite{Kazantsev91}). The temperature of the cooling of neutral atoms (use H and hydrogen-type atoms Na, Cs) has been reached up to the fraction of mK for relatively large samples of atoms and for minutes in a time that provided by engaging of atoms by magnetic trap with zero field at the centre of the system. The typical density $n$ was about $10^{11}$- -  $10^{12} \textrm{cm}^{-3}$. For such densities the quantum effects become essential close to the Bose condensation temperature
\begin{equation}
T_B = \frac{3.31 \hbar^2 n^{2/3}}{g^{3/2}m} \notag{}
\end{equation}
and below. For example, this temperature for H has the order of 1 mK for such density that is more or close to the achieved cooling temperature. For heavy atoms the reached temperature until yet is far from the corresponding $T_B$ but there exist some ideas how to get lower temperature \cite{SurdutovichPC}. Therefore one can hope that in the nearest years it will be get at the experiment the unique quantum gas.

The purpose of this paper is to treat the behaviour of such gas for the temperature less than the condensation temperature when almost all particles are in the condensate state (for simplicity we restrict ourselves by $T= 0$) when the radiation is absent. It should be noted that resonant electromagnetic wave absorption and forthcoming reemission lead to the repulsing between atoms of the Coulomb type \cite{Kazantsev91}. Without radiation the interaction between atoms in gas phase for long distance more than Bohr radius $a$ is defined by the Van der Waals attraction forces, $F= -\gamma/r^6$.  Then in the gas approximation following to \cite{Gross57} for oscillations of the condensate one can get the standard Hamiltonian for the nonlinear Schr\"{o}dinger equation (NLSE) \begin{equation}
\label{eq:Eqn2}
H = \int \left[\frac{\hbar^2}{2m}|\nabla \psi|^2+\frac{G}{2}|\psi|^4+\frac{\beta}{3}|\psi|^6\right]dr.
\end{equation}
Here $\psi$ is the wave function of the condensate, so that $n = |\psi|^2$ is the gas density. This constant $G$ is connected with the scattering length $\alpha = (m/2\pi \hbar)G$. The character of the interaction depends significantly on the $\alpha$ sign. For example, for hydrogen $\alpha = 1$ (in Bohr radii), for Cs $\alpha=100$, bu the sign is unknown. Therefore we consider both cases. In the repulsion case we neglect the next term of expansion over $n$ proportional to $\beta$. The main attention will be paid to the attraction because in opposite case there exists a lot of papers about, starting from the Bogolyubov famous paper. 
\section{Instability and Collapse}
We start our consideration from the attraction case. The equation corresponding to $H$ is the well-known NLSE, which we write in dimensionless variables for $G<0$:
\begin{equation}
i\psi_t = -1/2\Delta \psi-f(|\psi|^2)\psi
\end{equation}
where $f(|\psi|^2) = |\psi|^2-\beta|\psi|^4$. We shall assume that the initial state of the gas is homogeneous with constant density $n$ so that $\beta |\psi|^2\ll 1$. Because the theory of the NLSE is well known for this case \cite{Vlasov70, Zakharov86,Vlasov88, Malkin88, Zakharov89, Kolokolov73} we present here only the main facts from this theory paying the attention to the application to this concrete physical situation.

First of all let us  be reminded that such a state will be unstable with growth rate
\begin{equation}
\label{eq:Gamma}
\Gamma= k (nf'-k^2/4)^{1/2}.
\end{equation}
(Consider  $\psi = \psi_0 +U +iV$; $U, V \sim  \exp[rt+ikx]$.)
The threshold of this instability is defined from $f'=0$. It gives $n = 1/(2\beta)$, that can be considered as condensed phase density. If we start from gas we have the instability. The nonlinear stage of this instability will be the collapse, i.e., the formation of the singularity for finite time.

A few things should be noted about 3-D solitons as one possible variant of the instability development. The simplest solutions of this type $\psi = \phi \exp[i \lambda^2 t]$ represent the stationary point of $H$ for fixed number of particles $N = \int |  \psi|^2dr$:
\begin{equation}
-\lambda^2 \phi +\frac{1}{2}\Delta \phi+|\phi|^2\phi - \beta |\phi|^4\phi = 0.
\end{equation}
These solutions are unstable in gas phase according to the criterion: $dN/d\lambda^2 <0$ \cite{Kolokolov73}. The qualitative arguments of this instability can be computed from the scaling transformations, $\psi \rightarrow a^{-3/2} \psi(r/a)$ remaining $N$. As a result of $H$ becomes the function of parameter $a$:
\begin{equation}
H = I_1 /a^2 -I_2 /2a^3+\beta I_3/3a^6
\end{equation}
where $I_1 = \frac{1}{2}\int|\nabla \psi|^2 dr$, $I_2 = \int |\psi|^4 dr$, $I_3 = \int|\psi|^6 dr$. Maximum of this function corresponds to the unstable soliton as to the saddle point. Besides, this function for $\beta = 0$ occurs to be unbounded from below. It is the main reason of collapse in the system. From this point of view the collapse can be considered as the process of some ``particle" falling in an unbounded potential. If $\beta$ is finite then this process can be stopped for very large amplitude that corresponds to the minimum of this function, i.e., the soliton of the condensed phase. It should be noted that for $\beta = 0$ there is no stabilization or saturation of the instability. The compression will continue until the formation of a singularity. It follows strictly from the virial theorem:
\begin{equation}
\frac{d^2}{dt^2}\int r^2 |\psi|^2 dr < 4H.
\end{equation}
After twice integration we have that positive value $\int r^2 |\psi|^2 dr$ vanishes for finite time. Because for the initial state $H$ is negative after the formation of clusters with characteristic size $ l\propto k_{0}^{-1}$ the collapse becomes inevitable, where $k_0 = \sqrt{2n_0}$ corresponds to the maximum of $\Gamma$ from Eqn.~\eref{eq:Gamma}. Other questions to be raised concern the possible regimes of collapse. The first variant of the collapse is the so called strong regime of compression when all particles in the cavity occur in the singularity \cite{Zakharov86}. It is easy to understand that this process is a semi-classical one because from the usual quantum mechanics it is known that in the potentials $V \propto -r^{-\sigma}$ for $\sigma>2$ the falling becomes, near the centre, more classical. But such collapse is unstable with respect to short-wave perturbations resulting in the formation of weak singularities. For this type of collapse the captured number of pa!
 rticles is formally equal to zero, in practice it is quite small. The most simple argument as to why instead of adiabatic compression the weak singularity appears is the following. Let us consider some region $\Omega$ with negative Hamiltonian $H$ (we call this region a cavity). Then with the help of the mean value theorem the inequality
\begin{equation}
\textrm{max} |\psi|^2 \leq |H_\Omega| / 2 N_\Omega
\end{equation}
follows. Here $N_\Omega$ is the number of particles in $\Omega$. Let us imagine that from $\Omega$ we have radiation of waves (or emission of particles). Because these emitted particles carry out the positive portion of $H$ (strictly speaking it is valid for the separate cavity) there is a reduction in $H$ for $\Omega$, i.e., $H$ for this region becomes more negative. It is evident that $N_\Omega$ vanishes. So the ratio $|H_\Omega |/2N_\Omega$ tends to infinity together with max $|\psi|^2$. It means that the radiation of waves promotes the collapse, leads to its acceleration. The corresponding solution to this regime is the self-similar one of Eqn.\~(2) for $\beta = 0$:
\begin{equation}
\psi = (t_0 - t)^{-1/2-i\epsilon} g\left(r/(t_0 -t)^{1/2}\right)
\end{equation}
where $g(\xi)$ obeys the equation
\begin{equation}
-(i-2\epsilon)g - (\xi \nabla)g+\Delta g+2|g|^2 g = 0.
\end{equation}
Hence the characteristic radius of the collapsing solution is proportional to $(t_0-t)^{1/2}$. From this fact follows that particles that reach the critical density $n = 1/\beta$ will form drop with the radius $r_0 = \beta^{1/2}/|g(1)|$. The resulting number of particles in the drop can be estimated as 
\begin{equation}
N\approx |g(1)|^{-3} \beta^{1/2} 4\pi/3.
\end{equation}
Because of $\beta n\ll 1$ this number is less than one. It means that at the collapse time $t_0$ the formation of condensed phase does occur, secondly, the energy gathered by the condensate particles during the collapse is quite large and so there exists the unique possibility to transfer this energy (or a significant amount of it) to undercondensate particles radiated from the vicinity of the singularity. It is very important also that while weak collapse the attraction potential will be formed for other non-collapsing particles which can be trapped after $t_0$. It follows, in particular, from asymptotic of $g(\xi)$ for $\xi \rightarrow \infty$:
\begin{equation}
g(\xi) \rightarrow A/\xi^{1+2i\epsilon} \textrm{ or } |\psi|^2\rightarrow|A|^2r^{-2}.
\end{equation}
Because during the weak collapse the number of trapped particles into singularity is very small, it is natural to suppose that the post-collapse regime is quasi-stationary. In searching for its distribution we can simply find a stationary solution. It is convenient to rewrite Eqn.~\eref{eq:Eqn2} for density $n$ and phase $\Phi$. Then simple analysis gives the following solution \cite{Vlasov88, Malkin88, Zakharov89}:
\begin{align}
n(r)& \approx 1/2 \ln(r/r_0)r^{-2}; \\ 
v(r)& = \partial \Phi/\partial r\approx P/2\pi \ln(r/r_0).
\end{align}
Here $P$ is the particle flux to the centre of this black hole and $r_0$ the size of the growing drop. The flux, from one side, forms the drop and, from the other, the counter flux of undercondensate particles which carry out the energy of the order of the kinetic energy of the falling particles. It is easy to estimate the corresponding energy $E$ lost by the cavity during the total time of collapse: 
$E\propto |E_{\textrm{in}}|( \beta n_0)^{-1}$ 
where $E_{\textrm{in}} $ is the initial gas energy. Thus the carried energy appears to be quite a large value. It can be assumed that a sufficiently small portion of particles (at least, less than half of the falling particles) will be emitted. These particles will interact weakly with condensate particles. Their phases can be considered random and therefore their interaction will be described by a four-wave kinetic equation. As known \cite{Zakharov86}, this interaction leads to the formation of a power-type Kolmogorov spectrum in the particle energy space realizing the particle flux in this space, i.e., forms the particle condensate. Now we don't know what kind of equilibrium is realized in this case.

Now let us estimate the growth rate of the drop. If we assume that approximately half of particles from the cavity with the size $l \propto k_{0}^{-1}$ are absorbed into the singularity, i.e. $N_{dr} \propto n_0 l^3 \propto n_{0}^{-1/2}$, then the characteristic time of the drop growth has the same order as the inverse growth rate (3): $\Gamma_{\textrm{max}} N_{dr}\propto l^2 P$. Hence $P\propto n_{0}^{3/2}$. Substituting this expression in the definition of $P$ one gets the drop growth as $\dot{V}\propto \beta n_{0}^{3/2}$ where $V$ is the drop volume.
At the end of this part we present the concrete numbers for an atom of  Cs mass. For density $10^{11}\textrm{cm}^{-3}$ the modulation length $l\sim 10^{-3}$cm, the collapse time ~$0.3$ sec that less than sample cooling size and the particle life time in the trap, respectively. It means that the observations of the discussed above effects represent quite real, of course, if the corresponding temperatures can be reached.

\end{document}